\documentclass[conference]{IEEEtran}

\usepackage{fixltx2e}

\usepackage{graphicx}
\usepackage{graphics}

\begin{document}
%
\title{Towards Dependable Change Management and Traceability for Global Software Development}

\author{\IEEEauthorblockN{David Ebo Adjepon-Yamoah}
\IEEEauthorblockA{Centre for Software Reliability\\
School of Computing Science\\
Newcastle University\\
Newcastle-upon-Tyne NE1 7RU, UK\\
Email: \textit{d.e.adjepon-yamoah@ncl.ac.uk}}
}


%


\maketitle

\begin{abstract}
This paper reports on our definition of guidelines for managing global software development (GSD) that implements the specific practice - \textit{manage requirements changes} - of the \textit{Capability Maturity Model Integration} (CMMI) Level 2. The guidelines present a model for change management and traceability that supports the implementation of the specific CMMI Level 2 goal. Also, to support the effective management of the system engineering processes, an adaptation of the \textit{Project Management Body of Knowledge} (PMBOK) process group (PG) for project life-cycle practices is provided. We introduce a cloud-based \textit{Reactive Middleware} which provides services for managing GSD projects towards dependable change management and traceability.

\end{abstract}

\section{Introduction}\label{intro}
System (or software) engineering is increasingly being practised over distributed geographical locations. This practice is classified as global software development (GSD). The growth in information technology such as cloud computing enables different projects to be developed at geographically distributed sites \cite{6297161}. The acceptance of the GSD process is because of its different benefits such as low cost, good productivity, access to skilled work forces and access to market \cite{SPIP:SPIP417}. 

However, GSD is highly dependent on \textit{requirements management} to ensure that the distributed teams are meeting project requirements. Requirements management is a collaboration intensive activity, which produces and consumes numerous artefacts during a project's lifespan. Software requirements   continuously change during the software development phases and it becomes very difficult to manage these changed requirements \cite{6297161}. Developing a shared understanding and awareness of these artefacts is integral to software collaboration \cite{DBLP:books/daglib/p/WhiteheadMGH10}, a challenge that globally distributed teams face on a regular basis. Use of collaborative technologies and artefacts is pervasive in GSD to alleviate some of the barriers caused by distance (geographic, temporal, and cultural), and to facilitate consistency in understanding and managing requirements \cite{1687861}. 

Also, the management of system engineering across all phases relative to meeting system requirements has been identified as the main source of software project failures \cite{5232804}. According to the Chaos Report 2013 released by the Standish Group, among all the large IT application development projects in 2012, only 10\% of the them were classified as successful, while 52\% were identified as challenged (completed but failed to achieve project initial goals such as time, cost or quality requirement), and the rest 38\% of the projects totally failed (cancelled before completion or never implemented) \cite{5232804}. 

This however, calls for the adoption and integration of; (1) \textit{software process improvement models} such as \textit{capability maturity models} (e.g. CMMI \cite{ProductCMMIfor2010} and ISO/IEC 15504) that focus on change management and traceability, and (2) \textit{well established management practices} such as those presented by the \textit{Project Management Body of Knowledge} (PMBOK) \cite{5937011}. Such an approach is essential for managing software or system engineering projects in a way that does not unnecessarily burden project managers. 

We report on the definition of a set of \textit{guidelines for managing GSD} that implements the specific practice - \textit{manage requirements changes} - of CMMI Level 2. These guidelines are facilitated with a \textit{Reactive Middleware} which applies a model for change management and traceability (CM-T), that supports the implementation of the specific goal. Also, to support the effective management of the system engineering processes, an adaptation of the PMBOK process group (PG) for project life-cycle practices is provided. 

We aim to answer the research question \textbf{(RQ)}: \textit{"How can the reactive middleware guide system engineering to ensure the continual tight linkage of stakeholders' requirements and system engineering processes?"}, by validating the following hypotheses; \textbf{H1}: Changes in system engineering requirements-related artefacts are captured and adequately propagated to all the related system engineering processes and stakeholders, and 
\textbf{H2}: It is possible to trace system development artefacts from creation and through system engineering processes.

\section{Reactive Middleware}
The \textit{Reactive Middleware} (RM) introduces a structured \textit{role-based management} of system development artefacts. Role in this context is described as stakeholders' responsibility (i.e. \textit{privilege}) to system artefacts. The RM is \textit{"reactive"} due to its cloud-based services for \textit{change management} and \textit{traceability}. This approach assigns priority to system requirements relative to their \textit{importance}. Also, the approach uses six main privileges (with roles): \textit{None}:- Have no access to the system artefact(s) for \textbf{PAWNS}, \textit{View}:- Only sees the system artefact(s) for \textbf{PAWNS}, \textit{Modify}:- Can see (view) and change the system artefact(s) for \textbf{MODIFIERS}, \textit{Review}:- Can see and change a modification to system artefact(s), in response to a set of notifications for \textbf{REVIEWERS}, \textit{Create}:- Can create and modify (view, modify) system artefact(s) for \textbf{CREATORS}, and \textit{Own}:- Full access (view, modify, review, delete, recall) to the system artefact(s) for \textbf{TEAM LEADERS}. Here, \textit{recalled} artefacts are reinstated deleted artefacts. 

\begin{figure}[h!]
  \begin{center}
    \includegraphics[width=9cm, height=4.61cm]{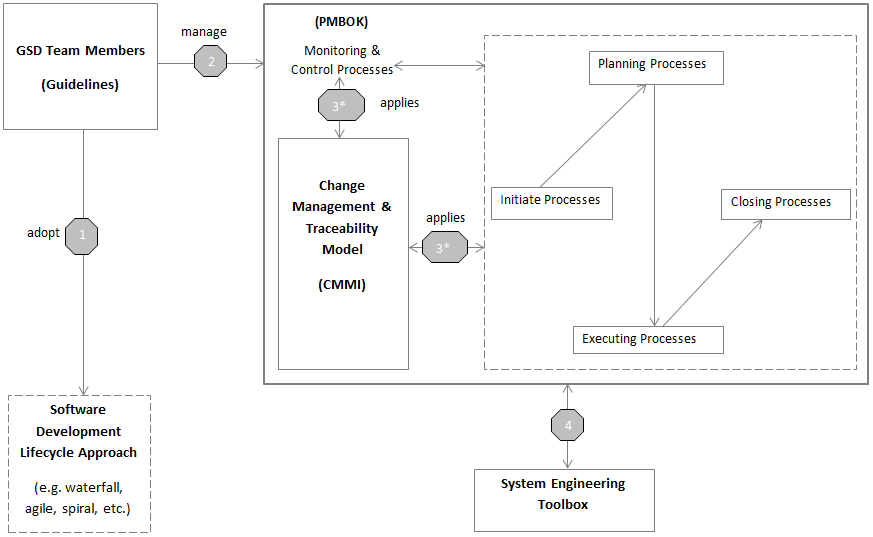}
    \caption{Reactive Middleware Interactions}
    \label{FigRMInteractions}
  \end{center}
\end{figure}

\begin{table}[h!]
\centering
\caption{Global Software Development Management Guidelines}
\label{tab:GSDGuide}
\resizebox{\columnwidth}{!}{%
    \begin{tabular}{ |p{1.1cm}| p{11.5cm}|}
    \hline
    \textbf{Guideline Steps ID} & \textbf{Guideline} \\ \hline 
    
    \textbf{GS1} & System development teams should appoint \textit{team leaders}. \\ \hline
     \textbf{GS2} & These team leaders will constitute the \textit{GSD change managers} \\ \hline
     \textbf{GS3} & System requirements classified based on identified \textit{dependability quality attributes} (i.e. safety, reliability, robustness, etc.), and are then \textit{prioritised} relative to their importance to the system stakeholders. \\ \hline
      \textbf{GS4} & Team leaders must assign roles to all team members with the prioritised requirements in mind, and manage the development process with the adapted PMBOK guide. \\ \hline
     \textbf{GS5} & All other \textit{change agents} especially the system engineering tools should be assigned a default privilege of \textit{review}. \\ \hline
     \textbf{GS6} & All system artefacts should be saved in a shared artefacts repository. \\ \hline
     \textbf{GS7} & The privileges (i.e. none, view, modify, review, own) of system stakeholders or change agents will determine the access privileges to system artefacts. \\ \hline 
    \textbf{GS8} & Change agents must \textit{subscribe} to relevant artefacts after they are created, in order to receive \textit{notifications} when they are changed. \\ \hline
    \textbf{GS9} & All \textit{related} artefacts must be linked together to facilitate traceability. \\ \hline 
    \textbf{GS10} & Changes made to any system artefacts must be \textit{logged}. \\ \hline
    \textbf{GS11} & When changes affect the high priority set of requirements, appropriate local team leader must lead the change request review process (i.e. involving the CM-T model) of the \textit{GSD change managers}. \\ \hline
    \textbf{GS12} & On the other hand, conflicts arising from changes to low priority set of requirements are resolved locally, lead by the local team leader. \\ \hline
    \textbf{GS13} & Changes in system artefacts should be traceable to manage its impact on related/linked requirements or artefacts. \\ \hline
    \end{tabular}
    }
\end{table}

\begin{figure}
  \begin{center}
      \includegraphics[width=9cm, height=6.8773cm]{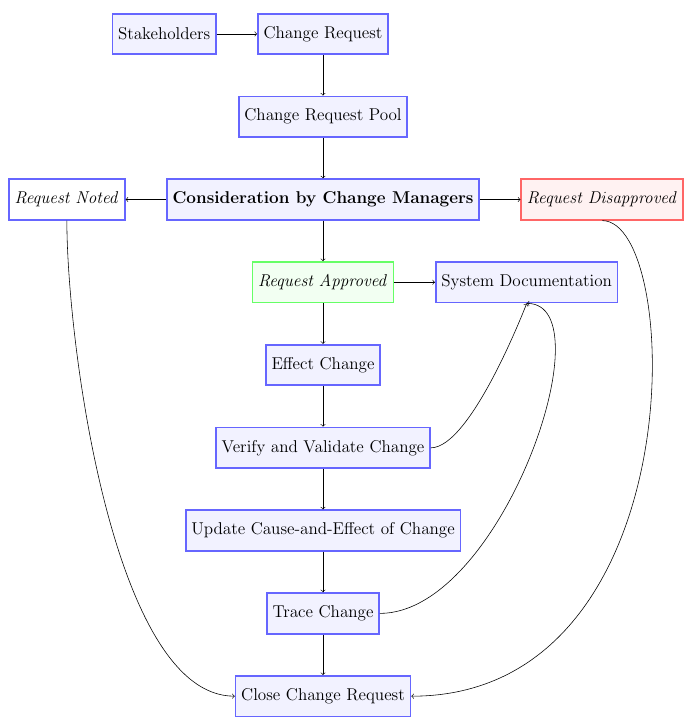}
    \caption{Change Management and Traceability Process Model}
    \label{FigCM-T}
  \end{center}
\end{figure}

The RM is composed of the \textit{Publish/Subscribe system} (PSS) and the \textit{Artefacts Monitoring system} (AMS). The PSS facilitates the subscription of system stakeholders to prioritised artefacts, to which they will initiate \textit{change requests} or will receive \textit{change notifications}. Also, the AMS is responsible for monitoring the relevant artefacts for changes, and then triggers the PSS to notify appropriate stakeholders or change agents. The RM interacts with the system stakeholders, \textit{System Engineering Tools}, and a \textit{Shared Artefacts Repository}. From Figure \ref{FigRMInteractions}, the GSD Team Members have the flexibility to \textit{adopt} any type of software development life-cycle (SDLC) approach (e.g. waterfall, agile, spiral, etc.) that suits their development style (i.e. \textit{Step 1}). Then following the prescribed \textit{management guidelines} (see Table \ref{tab:GSDGuide}) featured by the Reactive Middleware, the GSD Team Members \textit{manage} the development process with the PMBOK process group for system engineering life-cycle (i.e. \textit{Step 2}). When there are \textit{change requests} that are related to the high priority requirements, the GSD change managers \textit{apply} the CM-T process model (see Figure \ref{FigCM-T}) to either \textit{approve}, \textit{note} (i.e. to be applicable in the future) or \textit{disapprove} the request (i.e. \textit{Step 3*}). This CM-T model takes into consideration the traceability of the change agents (i.e. system stakeholders, artefacts and tools) involved in the change request. System engineering tools form an important change agent in the development process (i.e. \textit{Step 4}).

\section{Conclusions}\label{conc}
We report on our work which defines a set of management guidelines for GSD projects. We introduce a Reactive Middleware that applies a defined change management and traceability process model, and an adapted PMBOK quality process management approach to GSD. The CM-T process model is evaluated using \textit{expert feedback process}, whiles the management guidelines are evaluated using an \textit{airlock control system} case study. We then answer the research question \textbf{(RQ)}, by validating the hypotheses (\textbf{H1} and \textbf{H2}) presented earlier. 

%
%



\bibliographystyle{IEEEtran}
\bibliography{mybib}
%
%
%

\end{document}